\documentclass[reprint,showpacs,showkeys,prb,aps,noeqsecnum]{revtex4-1}

\usepackage{graphicx}
\usepackage{epstopdf}
\usepackage{bm}
\usepackage{amsmath}
\usepackage{subfigure}
\usepackage{mathrsfs}

\begin{document}

\title{On the existence conditions of surface spin wave modes in (Ga,Mn)As thin films
}

\author{H.~Puszkarski}
\email{Corresponding author: henpusz@amu.edu.pl}
\affiliation{Surface Physics Division \\ Faculty of Physics, Adam Mickiewicz University 
ul. Umultowska 85, 61-614 Pozna\'n, Poland} 
%	\address{}
\author{P.~Tomczak} 
 %       \email{email: ptomczak@amu.edu.pl}
	\affiliation{Quantum Physics Division\\ Faculty of Physics, Adam Mickiewicz University 
ul. Umultowska 85, 61-614 Pozna\'n, Poland}

%\maketitle

\date{\today}

\begin{abstract}

  Spin-wave resonance (SWR) is a newly emerged method for studying surface magnetic anisotropy and surface spin-wave modes (SSWMs) 
in (Ga,Mn)As thin films. The existence of SSWMs in (Ga,Mn)As thin films has recently been reported in the literature; 
SSWMs have been observed in the in-plane configuration (with variable azimuth angle $\varphi_M$ between the in-plane magnetization 
of the film and the surface [100] crystal axis), in the azimuth angle range between two in-plane \emph{critical angles} $\varphi_{c1}$ and $\varphi_{c2}$. 
We show here that cubic surface anisotropy is an essential factor determining the existence conditions of the above-mentioned SSWMs:  
conditions favorable for the occurrence of surface spin-wave modes in a (Ga,Mn)As thin film in the in-plane configuration 
are fulfilled for those azimuth orientations of the magnetization of the sample that lie around the hard axes of 
cubic magnetic anisotropy. This implies that a hard cubic anisotropy axis can be regarded in (Ga,Mn)As thin films as an easy axis for surface spin pinning. 

\end{abstract}


\maketitle

\section{Introduction}

Recent studies of the electronic structure of (Ga,Mn)As by spin-resolved photoemission \cite{sadowski} 
indicate that the surface region of a (Ga,Mn)As sample can be regarded 
as a layer with properties qualitatively different from those of the underlying bulk. 
In particular, the authors of the cited paper demonstrate that the surface of this material 
has a ferromagnetic character even at room temperature; 
consequently, in these temperature conditions a (Ga,Mn)As sample will include 
a ferromagnetic phase confined to a very thin surface region. 
The above-mentioned finding can be a precious contribution to the ongoing debate \cite{dobrowolska, samarth} 
on the essential mechanisms increasing the Curie temperature in (Ga,Mn)As, 
since in the light of the cited paper \cite{sadowski} 
in-depth studies of the nature of the surface ferromagnetism in this material 
could bring us closer to the understanding of the processes that might increase its Curie temperature even to the room temperature range.

The present study is devoted to spin-wave resonance (SWR), 
a very effective experimental method for the study of 
ferromagnetic properties of (Ga,Mn)As thin films, and their surface in particular. 
It allows full characterization of the \emph{surface magnetic anisotropy}, 
which, as we demonstrate here, provides the basis for the determination of 
the existence conditions of magnetic (spin-wave) modes localized at the \emph{surface} of a (Ga,Mn)As sample. 
We believe that this result is a step forward in exploring the differences 
in ferromagnetic properties between the surface and bulk of this material.

\section{SWR Surface Inhomogeneity Model}

The existence of surface spin-wave modes (SSWMs) in (Ga,Mn)As thin films was first reported by Liu {\it et al.} \cite{liu}, which observed 
SSWMs in both the out-of-plane configuration, with variable polar angle~$\vartheta_M$
between the magnetization of the film and its surface normal, and the in-plane configuration, with variable azimuth angle~$\varphi_M$ between 
the in-plane magnetization of the film and the [100] crystal axis. In the former case SSWMs only exist when the polar angle~$\vartheta_M$ is
larger than a certain angle~---~the {\it out-of-plane critical angle}~---~while in the in-plane configuration SSWMs are observed in the azimuth angle 
range {\it between} two {\it in-plane critical angle}s. 

These experimental findings are interpreted theoretically in our recent papers \cite{HP+PT14, HP+PT15, HP+PT15b}, 
in which we propose an appropriate model of surface anisotropy that both explains the existence of critical angles
and allows to determine the conditions of existence of SSWMs in full agreement with the experimental data. 
In our model of surface anisotropy we use the concept of surface pinning parameter~$A_{sur\!f}$, 
which describes the freedom of precessing surface spins in relation to the freedom of precessing bulk spins. 
We have determined the functions~$A_{sur\!f}(\vartheta_H)$ and 
$A_{sur\!f}(\varphi_M)$ describing the configuration dependence of the surface parameter in the out-of-plane and in-plane configurations, respectively.
The explicit formula for the in-plane surface parameter reads\cite{HP+PT15}:
\begin{eqnarray}\nonumber
 A_{sur\!f}(\varphi_M)=
 1+a_{iso}+
 a_{uni}\,\mbox{sin}^2\left(\varphi_M-45^\circ\right) +\\
 a_{cub}\left[\left(3+\mbox{cos}4\varphi_M\right)+
 \left(3+\mbox{cos}4\varphi_M\right)^4\right].
\label{A_surf}
\end{eqnarray}
A satisfactory interpretation of the experimental SWR spectra 
obtained in the in-plane configuration \cite{liu} is achieved with the surface parameter described 
by the series presented above in equation~(\ref{A_surf}) 
(found in an investigation discussed in detail in \cite{HP+PT15}) with the following values 
of the series coefficients $a_{iso}$, $a_{uni}$ and~$a_{cub}$:
\begin{equation}
 a_{iso}=0.1058;\, a_{uni}=0.027;\, a_{cub}=-0.0023.
 \label{a_values}
\end{equation}
The series coefficients in equation~(\ref{A_surf}) describe   the contributions of different surface anisotropy 
components to the surface spin pinning in (Ga,Mn)As thin films: 
$a_{iso}$ is the isotropic contribution to the pinning, $a_{uni}$ is related to the uniaxial
component, and $a_{cub}$ describes the pinning due to the cubic anisotropy.

In our model the critical angles are determined by the condition
\hbox{$A_{sur\!f}(\varphi_M)=1$}, and the configuration angles for which SSWMs exist by the condition
\hbox{$A_{sur\!f}(\varphi_M)>1$}. Based on these conditions, in the present paper we will discuss the impact 
of each of the surface anisotropy components appearing in Eq.~(\ref{A_surf}), and provide a physical interpretation of their respective roles 
in the generation of SSWMs. To this end we will use a surface pinning diagram built in the polar coordinate system; 
using this pinning diagram we will
obtain a graphical representation of the configuration dependence of the pinning felt by the surface spins.

\section{ Surface pinning diagram  vs. SWR spectra}

\begin{figure}[h]
\begin{center}
\includegraphics[width=0.3\textwidth, angle=-90]{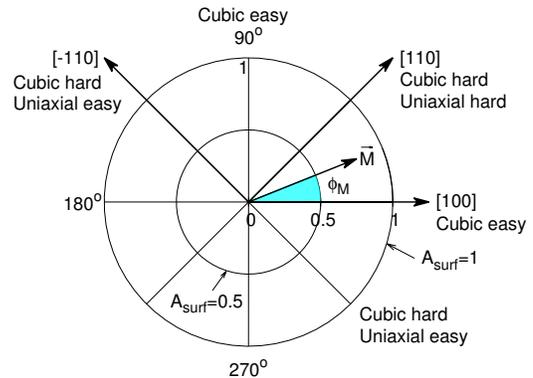}

\caption{The idea of \emph{surface pinning diagram}: each point in the plane represents the surface pinning conditions 
described by a specific value of the surface pinning parameter $A_{sur\!f}(\varphi_M)$ 
corresponding to the given magnetization direction $\varphi_M$. 
The circle $A_{sur\!f}(\varphi_M)=1$ corresponds to natural pinning conditions.\label{rys1}}

\end{center}
\end{figure}

The surface pinning diagram is a planar map showing how 
the dynamics of surface spins in a (Ga,Mn)As thin film changes with their orientation 
with respect to specific crystal axes that characterize the surface structure. 
The pinning diagram is based on a polar coordinate system lying in the plane of the film surface (see~Fig.~\ref{rys1}).
In this coordinate system the azimuth angle~$\varphi_M$ describes the orientation of the magnetization of the sample with respect to a reference axis, 
which is the [100] crystal axis; the distance between a point in the diagram and the pole of the coordinate system measures 
the value of the surface pinning parameter~$A_{sur\!f}=A_{sur\!f}(\varphi_M)$ corresponding to a given azimuth angle. Against the pinning 
diagram in Fig.~\ref{rys1} we have added the crystal axes characterizing the magnetic anisotropy in the (Ga,Mn)As sample used
in the spin-wave resonance~(SWR) study\cite{liu} (in which the SWR measurements were performed in the (001) in-plane configuration).

Introduced in papers \cite{HP70,HP79},
the concept of surface pinning parameter~$A$ was proposed to describe the degree of freedom of surface spins in their precession.
By definition the surface pinning parameter value~$A=1$ corresponds to the natural freedom of the surface spins resulting from breaking their
bonds with neighbors eliminated from the system by surface cut. This particular value of~$A$ divides the ($A_{sur\!f}(\varphi_M),\varphi_M$) plane
into two different regions: within the circle of radius~$A_{sur\!f}(\varphi_M)=1$, where only bulk spin-wave modes exist,
and region beyond this circle, in which SSWMs exist in the spectrum of allowed modes (see also \cite{LW, Urb, diep}). 

\begin{figure}[h]
\begin{center}
\includegraphics[width=0.25\textwidth, angle=-90]{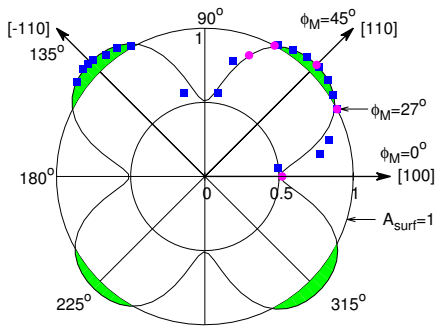}

\caption{Pinning trace, or the configuration dependence of the surface pinning, 
of a (Ga,Mn)As thin film plotted in the pinning diagram. 
The solid line represents the magnetization angle dependence 
of the in-plane surface pinning parameter resulting from our model (see Eqs (1-2)). 
This theoretical curve is found to fit very well the experimental data obtained by Liu~\emph{et~al.}~\cite{liu}; 
see the experimental points represented by squares.
\label{rys2}}
\end{center}
\end{figure}

The pinning diagram in Fig.~\ref{rys2} presents 
the configuration dependence of the surface parameter resulting from our formula (1) 
against the experimental data obtained by Liu~\emph{et~al.}~\cite{liu}; 
squares indicate the experimental surface parameter values corresponding to different configurations. 
The theory proves to fit very well the experiment. 
Moreover, as we will see below in Fig.~\ref{rys3}, 
the theory reproduces faithfully the experimentally observed evolution of the SWR spectrum with the azimuth angle~$\varphi_M$.

Figure~\ref{rys3} shows spin-wave mode profiles resulting from our theory 
and the corresponding SWR spectra for five representative in-plane configurations 
that correspond to azimuth angles~$\varphi_M=0^\circ, 27^\circ, 45^\circ, 62^\circ$ and $70^\circ$ 
(these configurations are also indicated in Fig.~\ref{rys2}, where they are marked with red circles). 
The most important feature of the configuration evolution 
that the experimental SWR spectrum undergoes as the azimuth angle increases to reach the consecutive values mentioned above 
is that from a multi-peak SWR spectrum at~$\varphi_M=0^\circ$ (with $A_{sur\!f}<1$) consisting exclusively of bulk modes, 
it becomes a single-peak FMR spectrum at the critical angle~$\varphi_{c1}=27^\circ$ (corresponding to $A_{sur\!f}=1$), 
and then, as the azimuth angle continues to grow above this critical value, 
regains its previous multi-peak form in the azimuth angle range $27^\circ<\varphi_M<62^\circ$, 
in which, however, it now includes a resonance peak corresponding to a surface mode 
(see the case with $\varphi_M=45^\circ$; $A_{sur\!f}(\varphi_M=45^\circ)>1)$. 
This result is a strong evidence that the SWR spectra measured in the in-plane configuration by Liu~\emph{et~al.}~\cite{liu} 
fulfill perfectly the assumptions of the Surface Inhomogeneity model, 
in which a key role is played by the surface magnetic anisotropy, constituting the source of the above-discussed effects.

\begin{figure}[h]
\begin{center}

\includegraphics[width=0.3\textwidth, angle=-90]{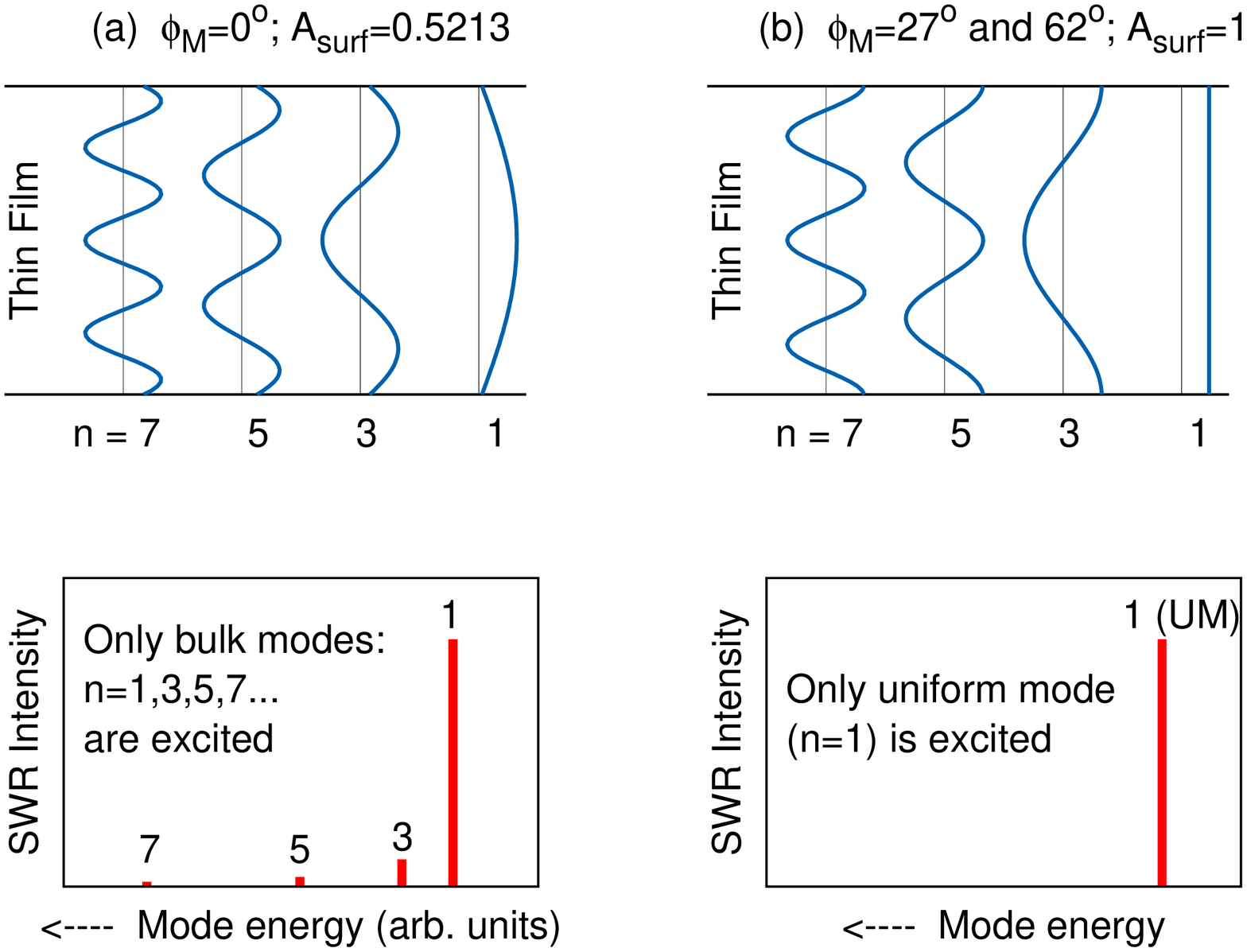}

\vspace{5mm}

\includegraphics[width=0.3\textwidth, angle=-90]{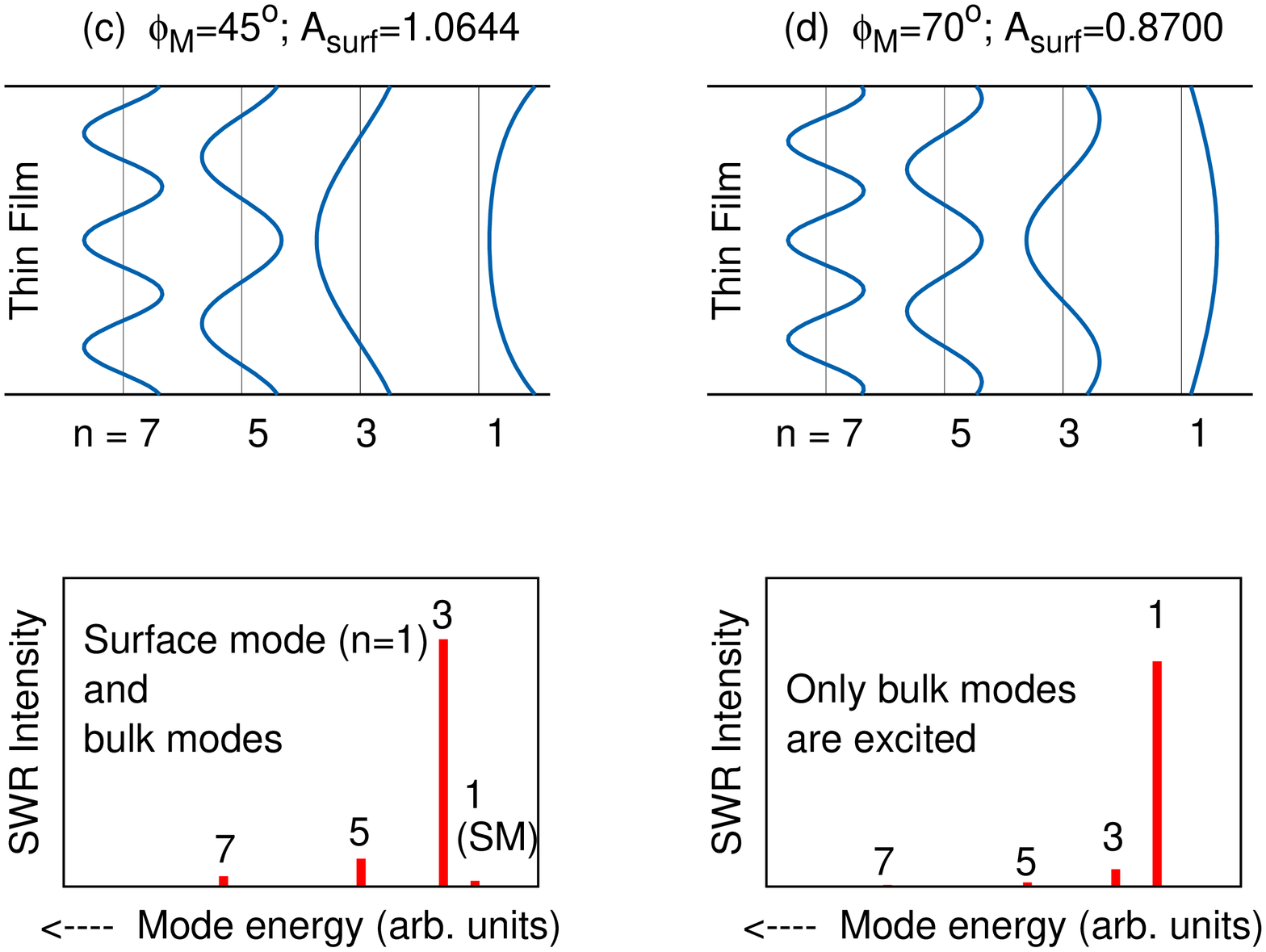}

\caption{Profiles of the lowest spin-wave resonance modes (top) and the corresponding SWR spectra (bottom) 
depicted separately for five in-plane configurations indicated by red points on the pinning trace in Fig.~\ref{rys2}. 
The spectra only exhibit peaks corresponding to symmetric modes of odd number, $n = 1, 3, 5, 7$. 
Two very peculiar effects are observed: firstly, the multi-peak SWR spectrum reduces to a single-peak FMR spectrum 
at two \emph{critical} azimuth angles, $\varphi_{c1}=27^\circ$ and $\varphi_{c1}=62^\circ$; 
secondly, in the angle range bounded by these two critical angles the spectrum includes a surface-localized resonance peak.
\label{rys3}}
\end{center}
\end{figure}

\section{Factors responsible for surface pinning and surface mode existence}

\begin{figure}[ht]
\begin{center}

\includegraphics[width=41mm, angle=-90]{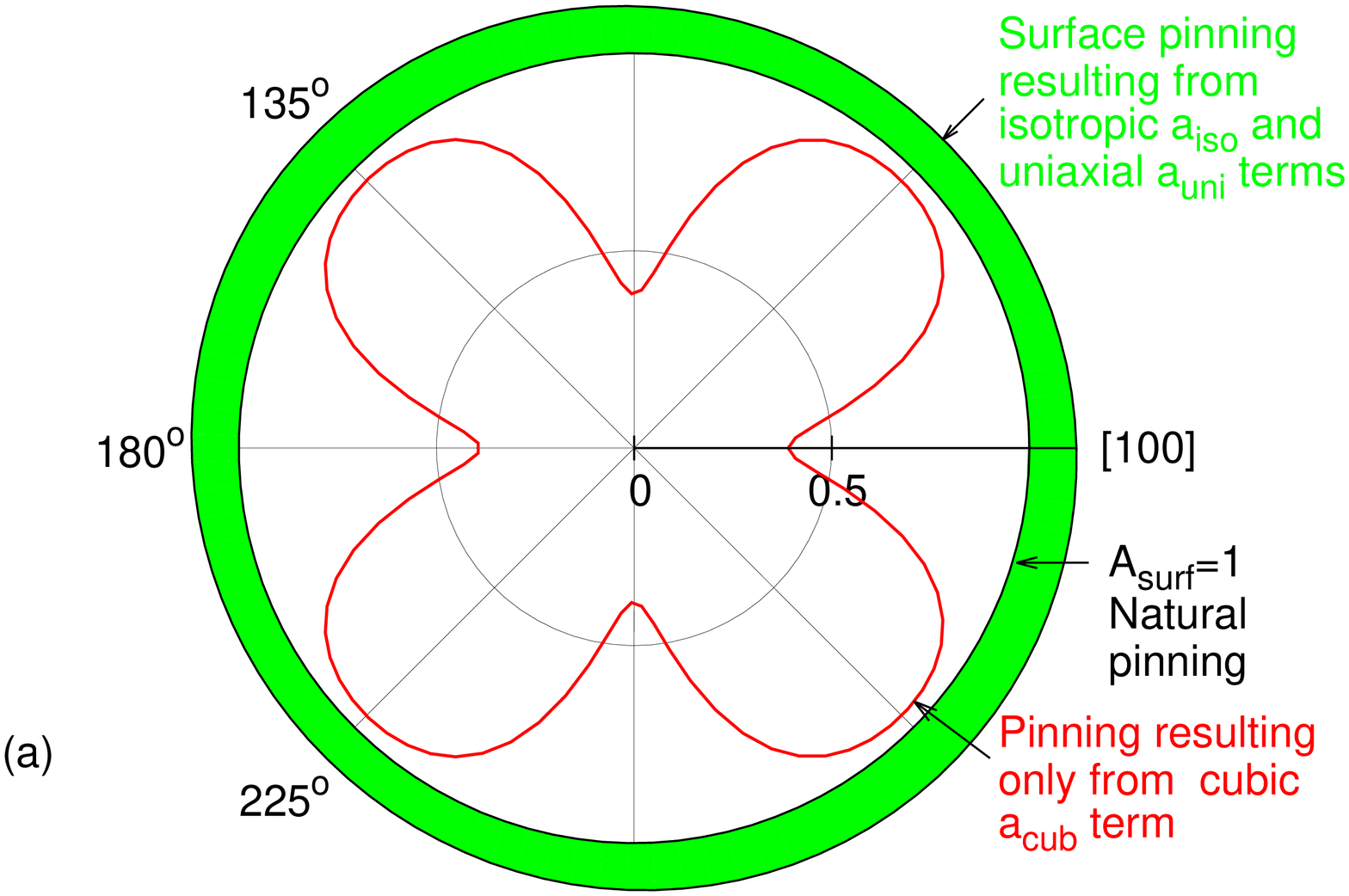}

\vspace{5mm}

\includegraphics[width=41mm, angle=-90]{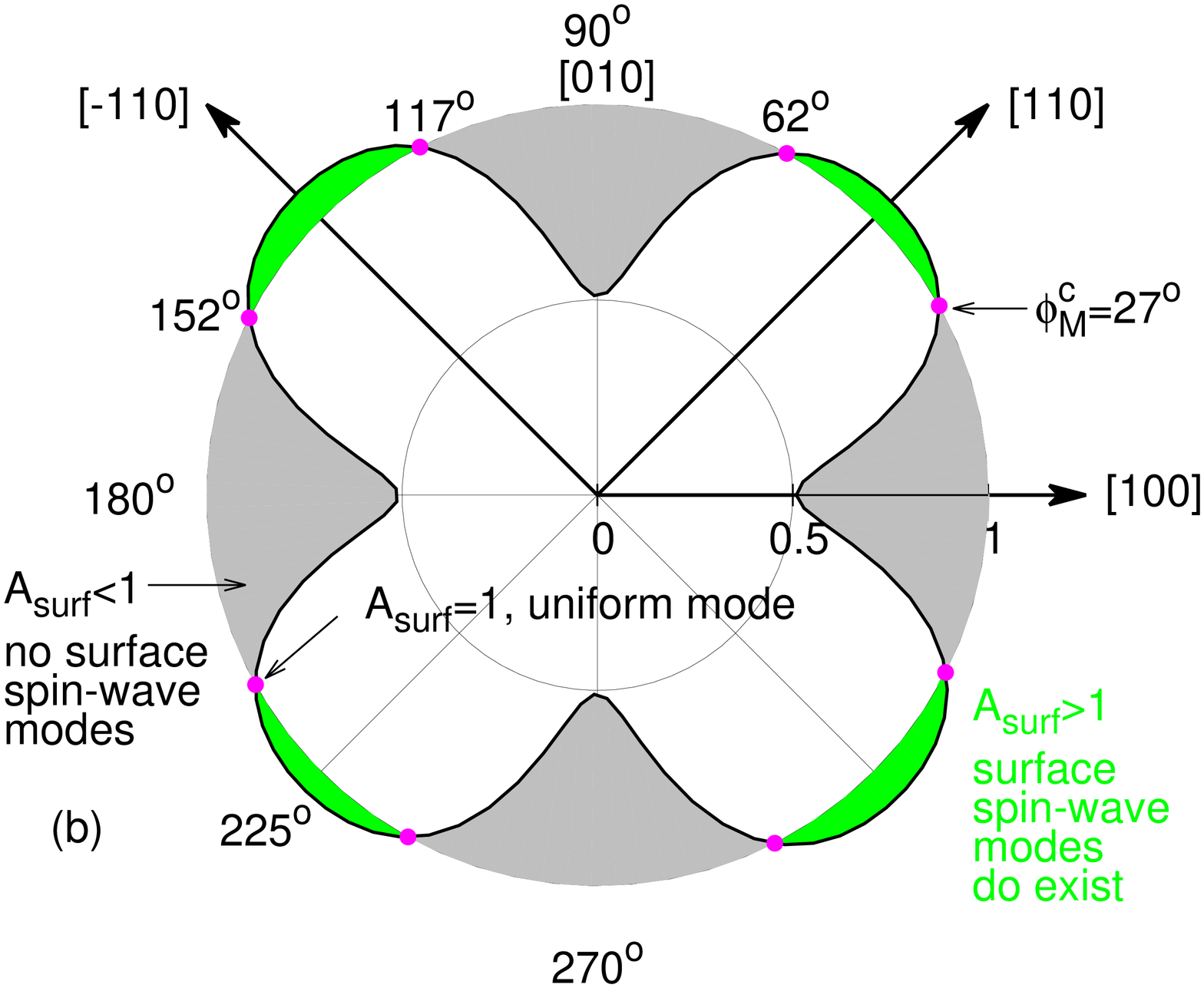}

\vspace{5mm}

\caption{Graphical analysis of the configuration evolution of the in-plane surface spin pinning 
in a (Ga,Mn)As thin film as described by Eqs (1-2). 
(a) Configuration dependence of two surface spin pinning \emph{contributions} resulting from 
\emph{different factors considered separately} (as indicated in the graph). 
(b) Represented by the solid line, the full surface pinning parameter $A_{sur\!f}(\varphi_M)$ 
as a function of the in-plane azimuth magnetization angle $\varphi_M$; 
regions of existence of SSWMs under these full pinning conditions are marked in green.
\label{rys4}}
\end{center}
\end{figure}

Figure~\ref{rys4}a shows two curves, each representing the azimuth angle dependence of a part of surface pinning
resulting from selected anisotropies -- the total contribution of the isotropic and uniaxial anisotropies 
is analyzed independently of the pinning related to the cubic anisotropy.
The first two contributions 
consonantly introduce the pinning to that region of the diagram in which SSWMs exist 
for any azimuth orientation (green ring in the diagram). In contrast, the pinning resulting 
from the cubic anisotropy, represented by a quadrifolium-like curve, is completely embedded in the region 
within the circle of radius~$A_{sur\!f}(\varphi_M)=1$; however, since
the circle~$A_{sur\!f}(\varphi_M)=1$ corresponds to the natural pinning, this implies that the cubic anisotropy alone
will not induce SSWMs in any azimuth orientation. Thus, already at this point, 
before proceeding to the analysis of the full pinning diagram presented in Fig.~\ref{rys4}b, 
we must realize that it will result from two above-mentioned opposite tendencies. 

As a result of the summation of the above-mentioned pinning contributions with different effects on the spins ~---~ one 
increasing their freedom, the other reducing it~---~in the full pinning diagram (Fig.~\ref{rys4}b) 
the green ring is reduced to four islands that still remain within the region
of surface mode existence. Each of these islands is associated with one hard cubic anisotropy axis, as indicated in Fig.~\ref{rys1}.
Also, each island corresponds to an azimuth angle range between the two critical angles for which $A_{sur\!f}(\varphi_M)=1$; 
each of these angle ranges surrounds symmetrically the cubic anisotropy
axis. \emph{It is in these azimuth angle ranges that SSWMs exist}. 
Beyond them the cubic anisotropy has a destructive effect on the surface modes;
this destruction is apparent in regions symmetric with respect to the \emph{easy cubic anisotropy axes}. 

Now we can provide physical grounds to these observations.
A hard magnetic axis is a direction in which spins are reluctant to align, 
and setting them in an equilibrium direction involves a high energy cost to the system.
Thus, the opposite tendency~---~divergence from the hard axis~---~must be more favorable energetically. 
This implies that a \emph{hard cubic anisotropy axis} plays the role of an \emph{easy axis} for the \emph{spin pinning}; 
we can refer to it as an \emph{easy pinning axis}, 
i.e., an axis from which spins diverge easily, since their pinning along its direction is weaker.
And vice versa: an \emph{easy cubic anisotropy axis} defines a direction that spins are reluctant to quit ~---~ a
\emph{hard surface pinning axis} that hampers the occurrence of SSWMs. Thus, the following final conclusion can be drawn from our considerations based on the pinning diagram:
\emph{in a (Ga,Mn)As thin film conditions favorable for the occurrence of surface spin-wave modes in the in-plane configuration
are fulfilled first of all for those azimuth orientations of the magnetization of the sample
that lie around the hard axes of cubic magnetic anisotropy}.
\vspace{3mm}
\section*{Acknowledgements}
This study is a part of a project financed by Narodowe Centrum Nauki (National Science Centre of Poland), Grant no. DEC-2013/08/M/ST3/00967.


\begin{thebibliography}{99}

\bibitem{sadowski} J.~Kanski, L.~Ilver, K.~Karlsson, M.~Leandersson, I.~Ulfat and J.~Sadowski,
{\it Electronic structure of (Ga,Mn)As revisited: an alternative view on the ``Battle of the bands''}, arXive: 1410.8842v2 [cond-mat.mtrl-sci].%[1]

\bibitem{dobrowolska} M.~Dobrowolska, K.~Tivakornsasithorn, X.~Liu, J.~K.~Furdyna, M.~Berciu, K.~M.~Yu and W.~Walukiewicz,
{\it Controlling the Curie temperature in (Ga,Mn)As through location of the Fermi level within the impurity band},
Nature~Materials {\bf 11}, 444 (2012).%[2]

\bibitem{samarth} N.~Samarth, {\it Battle of bands}, Nature~Materials {\bf 11}, 360 (2012).%[3]

\bibitem{liu} X.~Liu, Y.-Y.~Zhou, and J.~K.~Furdyna, {\it Angular dependence of spin-wave resonances and surface spin pinning 
in ferromagnetic (Ga,Mn)As films}, Phys.~Rev.~B~{\bf 75}, 195220 (2007).      % [4]

\bibitem{HP+PT14} H.~Puszkarski and P.~Tomczak, {\it Spin-Wave ResonancPhys. Status Solidi B,e Model of Surface Pinning in 
Ferromagnetic Semiconductor (Ga,Mn)As Thin Films}, Sci.~Rep.~{\bf 4}, 6135 (2014). %[5]

\bibitem{HP+PT15} H.~Puszkarski and P.~Tomczak, {\it Spin-Wave Resonance in (Ga,Mn)As Thin Films: Probing in-plane surface magnetic anisotropy}, 
Phys.~Rev.~B~{\bf 91}. 195437 (2015). %[6]

\bibitem{HP+PT15b} H.~Puszkarski and P.~Tomczak, {\it Model for the Surface Anisotropy Field Observed in Spin-Wave Resonance in (Ga,Mn)As Thin Films}, 
Acta~Phys.~Polon.~A~{\bf 127}, 508 (2015). %[7]

\bibitem{HP70} H.~Puszkarski, {\it Quantum Theory of Spin Wave Resonance in Thin Ferromagnetic Films. Part I: Spin Waves in Thin Films, 
Part II: Spin-Wave Resonance Spectrum}, Acta~Phys.~Polon.~A~{\bf 38}, 217 and 899 (1970).%[8]

\bibitem{HP79} H.~Puszkarski, {\it Theory of Surface States in Spin Wave Resonance}, Progr.~Surf.~Sci.~{\bf 9}, 191 (1979).%[9]

\bibitem{LW} L.~Wojtczak, {\it Magnetic~Thin~Films~(in~Polish)}, Wydawnictwo Uniwersytetu \L{}\'odzkiego, \L{}\'od\'z 2009. %[10]

\bibitem{Urb} A.~Urbaniak-Kucharczyk, {\it The Coherent Potential Approximation in the Description of Spin Wave Resonance},
Phys. Status Solidi B, ~{\bf 189}, 239 (1995).%[11]

\bibitem{diep} Diep-The-Hung and J.~C.~S.~L\'{e}vy, {\it Critical angle of external magnetic field for surface spin waves in thin ferromagnetic films}, 
Surface Science ~{\bf 80}, 512 (1979).%[12]

\end{thebibliography}
\end{document}